\documentclass[onecolumn,nofootinbib]{revtex4}
\usepackage{amsmath,amssymb,accents,graphicx}
\begin{document}

\newcommand{\myvec}[1]{\accentset{\rightharpoonup}{#1}}
\newcommand{\ket}[1]{| #1 \rangle}
\newcommand{\bra}[1]{\langle #1 |}
\newcommand{\op}[1]{{\mathbf #1}}
\newcommand{\ops}[1]{{\boldsymbol #1}}
\newcommand{\Mit}{\mathrm}
\newcommand{\Tr}[2][]{\mathrm{Tr}_{#1} \! \left[ #2 \right]}

\newcommand{\be}{${}^9\mbox{Be}^+$ \:}
\newcommand{\Sstate}{{}^2{\rm S}_{1/2} \:}
\newcommand{\Pone}{{}^2{\rm P}_{1/2} \:}
\newcommand{\Pthree}{{}^2{\rm P}_{3/2} \:}
\newcommand{\Plev}{\mathrm{P}}
\newcommand{\Fd}{F=2, m_F=-2}
\newcommand{\Fu}{F=1, m_F=-1}
\newcommand{\Fs}{{{\mathcal F}_\mathrm{S}}}
\newcommand{\Fone}{{{\mathcal F}_{1/2}}}
\newcommand{\Fthree}{{{\mathcal F}_{3/2}}}
\newcommand{\Fp}{{{\mathcal F}_\mathrm{P}}}
\newcommand{\Fk}{{{\mathcal F}_k}}
\newcommand{\Eps}{{\hat{\epsilon}}}
\newcommand{\phif}{\phi_\Mit{fluct}}

\newcommand{\ua}{{\uparrow}}
\newcommand{\da}{{\downarrow}}
\newcommand{\uan}{{\uparrow^{\mathrm{N}}}}
\newcommand{\dan}{{\downarrow^{\mathrm{N}}}}
\newcommand{\hc}{\mbox{h.c.}}
\newcommand{\cc}{\mbox{c.c.}}
\newcommand{\deltak}{\myvec{\Delta k}}
\newcommand{\omegaud}{\omega_{\da\ua}}
\newcommand{\omegafs}{\omega_{\rm FS}}
\newcommand{\sumj}{\sum_{j=1,2}}
\newcommand{\sumk}{\sum_{k=\{ 1/2,3/2 \}}}
\newcommand{\Omegaud}{\Omega_{\da\ua}}
\newcommand{\deltaud}{\delta_{\da\ua}}
\newcommand{\Oop}{\boldsymbol{\mathcal{O}}}
\newcommand{\nbar}{\overline{n}_\Mit{COM}}
\newcommand{\heat}{\Gamma_\Mit{heat}}

\newcommand{\mydash}{~$\leftrightarrow$~}

\newcommand{\antih}{$\overline{\mbox{H}}$ }
\newcommand{\htwo}{$\mbox{H}_2$}
\newcommand{\yb}{$\mbox{Yb}^+\:$}

\newcommand{\widtha}{14cm}
\setlength{\abovedisplayskip}{1mm} 

\title{Ion-trap quantum information processing: experimental status}

\author{D. Kielpinski}

\affiliation{Centre for Quantum Dynamics, Griffith University, Nathan QLD 4111, Australia}

\begin{abstract}

Atomic ions trapped in ultra-high vacuum form an especially well-understood and useful physical system for quantum information processing. They provide excellent shielding of quantum information from environmental noise, while strong, well-controlled laser interactions readily provide quantum logic gates. A number of basic quantum information protocols have been demonstrated with trapped ions. Much current work aims at the construction of large-scale ion-trap quantum computers using complex microfabricated trap arrays. Several groups are also actively pursuing quantum interfacing of trapped ions with photons.

\end{abstract}

\maketitle

\tableofcontents

\section{Introduction}
\label{sec-intro}

Quantum mechanics offers algorithms for efficient factorization of large numbers \cite{Shor-factoring-algorithm} and efficient searching of large databases \cite{Grover-search-algorithm}, two problems that appear insoluble in classical computing. Because much modern cryptography relies on the difficulty of factoring large numbers, a large-scale quantum computer could have a large impact on many areas of technology, Internet commerce being only one example. At the same time, the delicate and demanding nature of QIP, with every quantum accounted for, requires a more subtle technology than that used to construct a classical computer. The search for physical systems supporting QIP has ranged far and wide, across optics, atomic physics, and condensed-matter physics \cite{Nielsen-Chuang-QIP-book, Clark-expt-qc-CONF, Spiller-Kok-QIP-2005-rev}. Several physical implementations, namely linear optics, trapped ions, and superconducting electronic circuits, demonstrate the essential ingredients of QIP, including initialization to a known quantum state, efficient readout of the quantum state, long qubit coherence time, and universal quantum logic. In particular, small quantum computers have already been constructed with trapped ions \cite{Kielpinski-ion-trap-tutorial}, and a number of basic QIP algorithms \cite{Chiaverini-Wineland-quantum-fourier-xform}, quantum memory schemes \cite{Kielpinski-Wineland-DFS, Chiaverini-Wineland-QEC}, and communication protocols \cite{Barrett-Wineland-teleportation} have been demonstrated. In this review, we discuss the experimental status and prospects of ion-trap QIP, referring to the theory of QIP and to other physical QIP implementations only for context. This is not to underestimate the excellent work in these other areas, but only to keep the review at a manageable length. For a general overview of QIP, the reader is encouraged to consult Nielsen and Chuang's essential text \cite{Nielsen-Chuang-QIP-book}, and for a recent review, \cite{Spiller-Kok-QIP-2005-rev}. \\ \\

Most quantum information processing devices are made up of two-level systems, ``qubits,'' where each qubit is analogous to a single bit in a classical computer. The quantum state of the device encodes information, and an appropriate unitary evolution of the state of the register can perform a computing task. In our case, a qubit corresponds to a trapped ion, with the two qubit states being two electronic energy levels of the ion. Laser cooling of several trapped ions causes the ions to form a Coulomb crystal, in which the ions are held in the equilibrium positions given by the combination of the trapping force and their mutual Coulomb repulsion. The techniques of optical pumping and electron shelving, well known for decades, provide efficient initialization and readout. Trapped-ion qubits can have extremely long memory times, at least 20 seconds for optical \cite{Haffner-Blatt-robust-entanglement} and 10 minutes for microwave \cite{Bollinger-Wineland-10-minute-coherence} transitions from the ground state. The qubit energy levels are often identical to those used for the long-lived transitions of atomic clocks, and techniques originally developed for atomic frequency metrology have proved invaluable in the development of ion-trap QIP. \\ \\

Universal quantum computation requires both single-qubit operations and a nontrivial two-qubit logic gate. The key to two-ion logic is the collective motion of the trapped ions, which can be cooled to the quantum ground state and can store quantum information for hundreds of milliseconds \cite{Wineland-Meekhof-expt-issues-ion-QC, Turchette-Wineland-ion-heating}. Laser coupling of the ion spin states to this shared quantum degree of freedom allows engineering of indirect spin-spin interactions and thus two-qubit logic gates. These techniques have enabled the demonstration of many basic QIP protocols with trapped ions \cite{Gulde-Blatt-deutsch-josza, Barrett-Wineland-teleportation, Riebe-Blatt-teleportation, Schaetz-Wineland-dense-coding, Chiaverini-Wineland-quantum-fourier-xform, Brickman-Monroe-ion-grover-search, Reichle-Wineland-entanglement-purification}. \\ \\

Imperfect control of the quantum information processor leads to {\it decoherence}, the loss of information through correlation with an uncontrolled environment. In ion-trap QIP, decoherence arises mainly from environmental electromagnetic noise, from technical noise in the logic laser, and from spontaneous emission from excited electronics levels. A decoherence-free subspace (DFS) encoding has proved resistant to the main source of environmental noise \cite{Kielpinski-Wineland-DFS}. The complementary technique of quantum error correction has recently been demonstrated in principle \cite{Chiaverini-Wineland-QEC}, but repeated error correction, which offers the prospect of unlimited coherence time, is still only a dream. Operation of a large-scale quantum computer will probably require both quantum error correction and DFS encoding. \\ \\

No physical quantum information processor has yet demonstrated computational power anywhere near that of present-day desktop computers. A viable QIP implementation must therefore be {\it scalable}. In other words, the present-day techniques for QIP in a given physical system must remain useful for much larger versions of the same system without consuming a disproportionate amount of physical or computational resources. The most widely considered roadmap for large-scale ion-trap QIP builds on the successes of small ion-trap QIP devices by proposing a modular architecture of interconnected quantum registers \cite{Wineland-Meekhof-expt-issues-ion-QC, Kielpinski-Wineland-QCCD}. This roadmap is now being realised by investigators around the world and steps along the way have already proven crucial for demonstrations of several basic QIP protocols. \\ \\

The growth of interest in QIP since the discovery of Shor's factoring algorithm has fostered work on QIP implementations in a wide variety of atomic, optical, and condensed-matter systems (see \cite{Spiller-Kok-QIP-2005-rev} for a recent review). There is broad consensus on the requirements for QIP: efficient initialisation and detection of qubit states, long qubit coherence times, and a set of one- and two-qubit gates that enable access to the entire Hilbert space of the quantum register. Among QIP implementations, only trapped ions, superconducting circuits, and single-photon linear optics currently meet these criteria. Trapped ions make nearly ideal small quantum registers, but significant technical work is still required for large-scale quantum computing. It is easier to see how to fabricate a large number of superconducting qubits, but decoherence times are still very short. Single photons again have long coherence time, but no deterministic two-qubit gate has yet been demonstrated, so long chains of gates cannot be reliably implemented. While other physical systems may soon achieve the criteria for QIP, it is encouraging that large-scale QIP already appears feasible (if challenging) in these three cases.\footnote{A warning to novices: the intense experimental competition in this field has led some investigators to make exaggerated claims on the basis of their results, and these claims have sometimes appeared in reputable journals. I do not believe that this paper cites any work of that kind.} \\ \\

\section{Ion traps for QIP}
\label{sec-trap}

Almost all experiments in ion-trap QIP so far have used radio-frequency (RF) traps to confine ions under ultra-high vacuum. In these traps, one applies large RF voltages to an electrode structure made of conducting material in order to create a quadrupole electric field with a minimum in free space. For RF voltages of a few hundred volts and an appropriate frequency $\Omega_\Mit{RF}$ in the MHz range, the RF field induces a deep (few eV) ponderomotive potential that confines ions harmonically at the field minimum with an oscillator frequency $\nu \ll \Omega_\Mit{RF}$ \cite{ghosh}. In the QIP context, the residual ion motion at the RF drive frequency, or ``micromotion,'' can generally be neglected once the effects of stray electric fields have been cancelled. The low pressure ($\sim 10^{-14}$ bar) in the vacuum chamber means that collisions of residual gas atoms with the ions are infrequent, on the order of one every 100 seconds. The ions are then well isolated from environmental perturbations, enabling the precise quantum state control needed for QIP experiments. \\ \\

One popular method for producing ions uses a resistively heated oven to generate an atomic beam of the species to be ionised. The atomic beam passes through the trapping region and intersects a beam of low-energy ($<100$ eV) electrons emitted from a resistively heated thoriated tungsten filament. Electron-impact ionisation produces ions in the trapping region, and the strong, pseudo-conservative trapping force is sufficient for confining the hot ion sample, which can have a temperature of several thousand kelvin. Some experiments have replaced the oven-generated atomic beam by laser ablation of a solid target containing the species to be ionised \cite{Hendricks-Drewsen-all-optical-loading, Leibrandt-Chuang-surface-ablation-loading}. In many cases one can replace electron-impact ionisation by photoionisation \cite{Kjaergaard-Drewsen-photoionization-loading}, which offers the advantage of much higher ionisation efficiency and isotope selectivity. The latter property is particularly useful for species like \yb \cite{Balzer-Wunderlich-YbII-trap}, for which several isotopes have high natural abundance. \\ \\

\begin{figure}
\begin{center}
\includegraphics*[width=8.3cm]{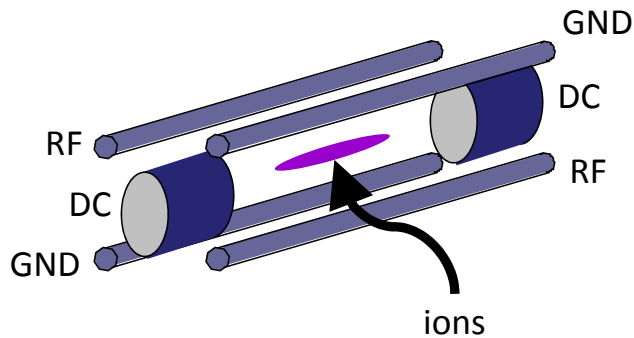}
\caption{Electrode structure of a linear RF trap. High-voltage RF applied to the rod electrodes provides ponderomotive confinement of ions to the trap axis, while the endcaps are held at DC potential for axial confinement. GND: electrical ground. The electrodes marked RF have equal RF voltages at all times. The DC electrodes are held at RF ground. }
\label{lintrap}
\end{center}
\end{figure}

Linear RF traps are preferred for QIP experiments because they nearly eliminate micromotion along one motional axis. A simple electrode structure for a linear trap \cite{Prestage-Maleki-linear-trap, Raizen-Wineland-linear-trap} is shown in Fig.~\ref{lintrap}. Essentially the trap is a quadrupole mass filter plugged at the ends with static potentials. To operate the trap, one applies RF voltage to the rods 1 and 3 of Fig.~\ref{lintrap}, while rods 2 and 4 are held at RF ground. The induced ponderomotive potential confines the ions to the RF nodal line, which lies along the $\hat{z}$ axis. The electrodes on the trap axis are held at a positive DC voltage relative to the  rod electrodes, pushing the (positive) ions toward the center of the trap. For motional amplitudes characteristic of laser-cooled ions, the resulting trap potential is harmonic in all three directions, with trap frequencies up to 50 MHz for light ions such as $\mbox{Be}^+$ \cite{Jefferts-Wineland-coax-trap}. Because the RF electric fields at the electrode surfaces are so high, electrical breakdown limits any attempt to radically increase the ion motional frequency \cite{Kielpinski-Wineland-thesis}. \\ \\

In traditional ion-trap experiments, the trap electrodes are fabricated from bulk metal and the electrode assembly is clamped into a custom jig, usually of ceramic. Since the most widely accepted path to large-scale QIP with ion traps requires the fabrication of large arrays of interconnected traps \cite{Wineland-Meekhof-expt-issues-ion-QC, Kielpinski-Wineland-QCCD}, recent experiments have moved toward microfabricated electrodes and quasi-planar (multilayer or single-layer) geometries. The first microfabricated traps used laser-machined substrates, but traps based on gallium arsenide \cite{Stick-Monroe-GaAs-ion-trap} and silicon \cite{Wineland-Seidelin-silicon-trap} have now been fabricated by MEMS techniques. The design and fabrication of trap arrays has been greatly simplified by the introduction of the ``surface'' trap geometry \cite{Chiaverini-Wineland-surface-trap-design}. In this design all electrodes lie in a plane, and a trap array of great complexity can be fabricated on an insulating substrate by a single photolithographic step. Surface traps using a printed-circuit-board substrate \cite{Brown-Chuang-PCB-ion-trap} allow rapid prototyping of designs for trap arrays. Silicon-based surface traps offer the possibility of very large scale integration of such a trap array with CMOS control circuitry \cite{Kim-Slusher-ion-trap-system-design}. \\ \\

\begin{figure}
\includegraphics*[width=8.3cm]{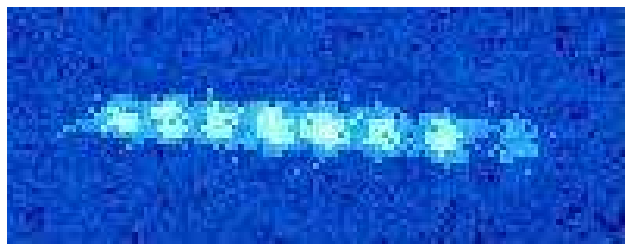}
\caption{A crystal of eight \yb ions in the trap at Griffith University, Brisbane, Australia. The ions are cooled by a 369.5 nm diode laser \cite{Kielpinski-Kaertner-diode-YbII-cooling} and are viewed by fluorescence imaging on a CCD camera. }
\label{eightions}
\end{figure}

When multiple ions are present in the trap, one must consider the Coulomb repulsion between ions as well as the ions' interaction with the harmonic trapping potential. If the ions are sufficiently cold, the classical equilibrium positions of the ions are given by minimizing the potential energy. For sufficiently weak axial confinement, the equilibrium positions all lie on the trap axis $x=y=0$, so that the ions line up in a string, as shown in Fig.~\ref{eightions}. Minimizing the combined trap and Coulomb potential gives the equilibrium positions along the axis \cite{Steane-ion-QIP-rev, James-trapped-ion-dynamics, Kielpinski-Wineland-sympathetic-cooling}. In the case of a two-ion string, the ion-ion distance is given by $(e^2/(2\pi \epsilon_0 m \nu_z^2))^{1/3}$, where $m$ is the ion mass and $\nu_z$ is the axial oscillation frequency of a single ion. In current experiments the separation between ions is on the order of 3 to 30 $\mu\mbox{m}$, with smaller separation for lower mass and for higher trap frequency. Hence it is experimentally difficult to maintain the positioning of a focused laser beam to the tolerance required for individual laser addressing of the trapped ions. While most current ion-trap QIP experiments use readout and quantum logic schemes that are designed to avoid the requirement of individual laser addressing, the Innsbruck group has had remarkable success in ion QIP using individual addressing (\cite{Nagerl-Blatt-individual-addressing}, and see below). \\ \\

After laser cooling, the residual motion of the ions is very small, so we can linearize the total potential about the equilibrium positions. The resulting harmonic oscillations constitute normal modes of the ion crystal. Solving the relevant eigenvalue equation gives the normal mode frequencies and eigenvectors (for numerical values see \cite{James-trapped-ion-dynamics}). For any number of ions, the lowest-frequency mode is always the center-of-mass (COM) mode, in which the ion string moves as a unit, with no relative motion between the ions. The Coulomb interaction then has no effect on the dynamics of the COM mode, so the COM frequency is just equal to the single-ion trap frequency $\nu_z$. Because of the symmetry of the ion string about $z=0$, the ions' relative amplitudes of motion in a given mode are either symmetric or antisymmetric about the center of the string. \\ \\

\section{Small ion-trap quantum registers}

\begin{figure}
\begin{center}
\includegraphics*{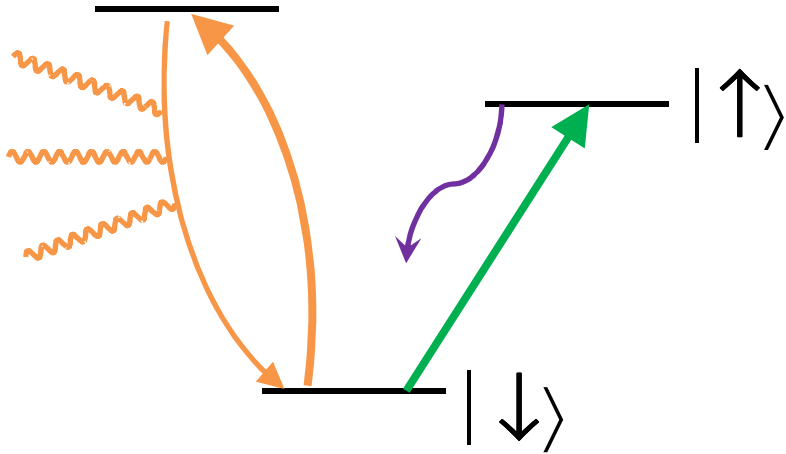}
\caption{Atomic energy levels for a trapped-ion qubit. The qubit transition between $\ket{\da}$ and $\ket{\ua}$ is shown in green. To detect the qubit state, one uses a laser to repeatedly drive the transition shown in orange, causing fluorescence from the $\ket{\da}$ state but not from the $\ket{\ua}$ state. The detection laser also cools the ions to the crystalline state. One reinitialises the qubit by driving a transition from $\ket{\ua}$ to an auxiliary energy level that decays into $\ket{\da}$. The auxiliary level can be the upper state of the detection transition or some other excited state. The purple arrow schematically indicates this reinitialisation process. }
\label{levelschem}
\end{center}
\end{figure}

Current experiments in trapped-ion QIP rely on laser excitation of electronic transitions for all QIP operations. Fig.~\ref{levelschem} shows a typical set of atomic energy levels for an ion that is being used as a qubit. There are two relevant transitions out of the electronic ground state $\ket{\da}$. One of these is a strong optical transition whose upper state $\ket{e}$ decays rapidly by spontaneous emission. This transition is used for initialisation and detection of the qubit. The other is a coherent optical or RF transition whose upper state $\ket{\ua}$ is extremely long-lived. The states $\ket{\da}$, $\ket{\ua}$ form the logical basis, corresponding to the role of 0 and 1 in a classical computer. In analogy to the spin-1/2 system, one often refers to the logical degree of freedom as the ``spin''. Coherent rotations of this spin correspond to single-qubit logic gates. Of course, all real ions have much more complex level structures than shown in Fig.~\ref{levelschem}, and so there are many variations and complications of this basic scheme depending on the exact level structure of the ion being used, but in almost all cases, the overall goal of these schemes is to reproduce the level structure of Fig.~\ref{levelschem}. \\ \\

\subsection{Readout and initialisation of ion qubits}
\label{sec-init}

Quantum computation requires the preparation of the computational register in a well-defined input state at the beginning of the computation and the efficient readout of the state of the register at the end of the computation. These are accomplished with the assistance of the strongly-allowed $\ket{\da} \rightarrow \ket{e}$ transition of Fig.~\ref{levelschem}, referred to as the readout transition. By choosing the correct ion species with an appropriate energy level structure, one arranges that the readout laser only induces fluorescence from $\ket{\da}$, while the laser is far off resonance for $\ket{\ua}$. Then a high fluorescence rate indicates $\ket{\da}$, and a low one, $\ket{\ua}$. This technique for high-efficiency internal state discrimination has been used in studies of trapped ions for decades \cite{Dehmelt-shelving-proposal, Wineland-Drullinger-electron-shelving, Sauter-Toschek-quantum-jump}. \footnote{Recently another efficient detection method has been demonstrated when the energy levels are not appropriate for the electron shelving technique \cite{McDonnell-Steane-Zeeman-detection}.} The fluorescence of the $\ket{\da}$ state distinguishes it from the $\ket{\ua}$ state with $>98\%$ efficiency per ion in each repetition of the experiment. The fluorescence has a total power of picowatts per ion and is emitted nearly isotropically, so current experiments in ion QIP use complex multi-element objective lenses to achieve large numerical aperture (NA) and high light collection efficiency. For current optical imaging systems, with f-number $\sim f/1$, the readout time can be as short as a few hundred microseconds \cite{Rowe-Wineland-Bell-inequality, Acton-Monroe-ion-readout-efficiency}. A recent experiment reports readout error rates as low as $1.5 \times 10^{-4}$ \cite{Myerson-Lucas-efficient-detection-PREPRINT}. Optical pumping on the same transition initialises the register with a residual error that is much smaller than the detection error and requires only a few microseconds. These excellent readout and initialisation properties are key ingredients in the success of ion-trap quantum computing. \\ \\

\begin{figure}
\begin{center}
\includegraphics*[width=8.3cm]{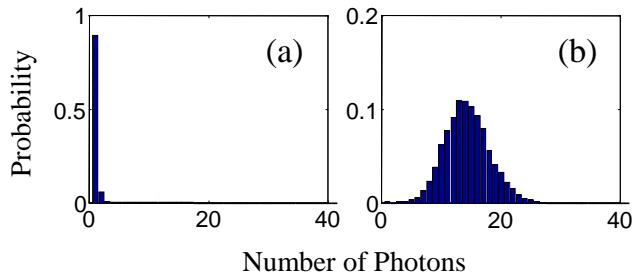}
\caption{Statistics of detected photons for a single ion prepared in a) the $\ket{\ua}$ state and b) the $\ket{\da}$ state. Statistics were collected over 1000 repetitions of the experiment. Figure courtesy NIST Ion Storage Group.}
\label{onehist}
\end{center}
\end{figure}

Qubit readout requires us to convert an analog quantity (number of photons measured) to a digital quantity (qubit state), so a knowledge of the photon statistics is crucial for high readout efficiency. Fig.~\ref{onehist} shows a histogram of the photon statistics for a single \be ion prepared in $\ket{\ua}$, and another histogram for an ion prepared in $\ket{\da}$. While the $\ket{\da}$ histogram is Poissonian, the $\ket{\ua}$ distribution has a long, non-Poissonian tail. The tail arises because one generally prolongs the detection period until off-resonant repumping of the $\ket{\ua}$ state to $\ket{\da}$ begins to cause photon scattering as well \cite{King-Wineland-THESIS, Acton-Monroe-ion-readout-efficiency}. The two histograms are readily distinguished, and by counting each readout of $>1$ photons as $\ket{\da}$ and all other readouts as $\ket{\ua}$, one can easily discriminate $\ket{\da}$ from $\ket{\ua}$ in a single repetition of the experiment with 98\% efficiency. Recent work has investigated the adaptive readout of an ion qubit over several repetitions of an experiment, reaching detection errors as low as $10^{-4}$, with a requirement of seven repetitions (on average) per readout \cite{Hume-Wineland-adaptive-qubit-meast}. \\ \\

\begin{figure}
\begin{center}
\includegraphics*[width=8.3cm]{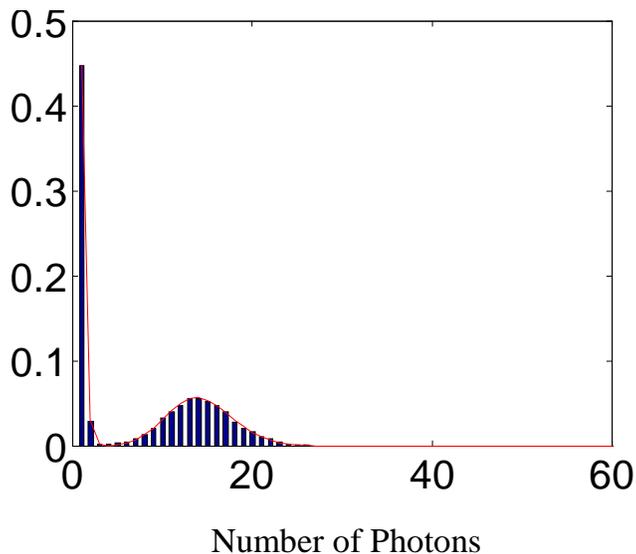}
\caption{Statistics of detected photons for a single ion prepared in the superposition state $\ket{\da} + \ket{\ua}$. Statistics were collected over 1000 repetitions of the experiment. The solid line is the best fit to the theoretical count distribution; the integral under the peak near zero photons is approximately equal to that under the peak near 15 photons, indicating nearly equal populations in the two states. Figure courtesy NIST Ion Storage Group.}
\label{superhist}
\end{center}
\end{figure}

In experiments with one ion, one might detect the ion in $\ket{\da}$ on some repetitions and in $\ket{\ua}$ on others, for instance by preparing the ion in a superposition state. Fig.~\ref{superhist} shows the photon statistics for the state $\ket{\da} + \ket{\ua}$. In each repetition of this experiment, the ion is projected into either the state $\ket{\da}$ or the state $\ket{\ua}$, in accordance with the quantum measurement postulate. Thus the probability distribution of photon counts is a linear combination of that found for $\ket{\da}$ (Fig.~\ref{onehist}a) and that found for $\ket{\ua}$ (Fig.~\ref{onehist}b), a direct verification of ``wavefunction collapse'' into an eigenstate of the measurement basis. \\ \\

For multiple ions, the simplest readout method is to collect the fluorescence from the entire ion string at once, making no attempt to spatially resolve the ions \cite{Kielpinski-ion-trap-tutorial}. In this case one reads out $N_{\ket{\da}}$, the number of ions in $\ket{\da}$, rather than independently reading out each qubit. The photon statistics over many repetitions of the experiment is compared with theoretical distributions that depend on the mean number of photons collected per ion in $\ket{\da}$, the mean number of photons due to background light, and the effect of off-resonant repumping \cite{King-Wineland-THESIS, Kielpinski-ion-trap-tutorial}. This method can resolve the relative probability of the outcomes $N_{\ket{\da}}$ for four ions with 98\% accuracy over 1000 repetitions and has been used successfully in six-ion experiments \cite{Leibfried-Wineland-six-atom-entanglement}. \\ \\

Even if the ions are not spatially resolved during detection, individual laser addressing of the ions can provide individual readout, though at the cost of considerable technical difficulty. Only the Innsbruck experiment routinely achieves individual addressing \cite{Nagerl-Blatt-individual-addressing}, and only at the laser wavelength used for quantum logic operations, rather than that used for detection. In this technique \cite{Riebe-Blatt-teleportation}, one performs a SWAP operation between $\ket{\da}$ and an auxiliary Zeeman sublevel of the metastable state $\ket{\ua}$ on all the ions except the one to be read out. Since all Zeeman sublevels of $\ket{\ua}$ are equally dark to the detection laser, fluorescence will be detected from the ion string only if the ion to be read out is in the state $\ket{\da}$. Successive SWAP operations and detection periods then read out the ions individually. \\ \\

\subsection{Single-qubit operations}

Universal quantum logic requires a set of single-qubit operations that are adequately realized by coherent coupling of the ion internal states \cite{Nielsen-Chuang-QIP-book}. A source of coherent radiation can drive the appropriate coupling if its frequency matches the energy splitting of the qubit states. Qubits using the hyperfine levels of the ground state (``hyperfine qubits'') are relatively easy to manipulate using commercial sources of microwave radiation, which easily achieve frequency stability on the 10 Hz scale. On the other hand, logic on qubits that use long-lived optical transitions (``optical qubits'') requires an ultrastable laser (less than 1 kHz linewidth) that can be tuned to the qubit transition wavelength \cite{Roos-Blatt-optical-qubit-control}. Such lasers present considerable technical challenges and are rarely found outside laboratories devoted to optical frequency metrology. However, the underlying physics of the atomic interaction with the radiation field is the same for both hyperfine and optical qubits. The same interaction again underlies nuclear magnetic resonance, where the relevant basis states are often the spin states of a proton in a magnetic field, thus the term ``spin'' for the qubit state. \\ \\

In manipulating an optical qubit, one attempts to achieve the ideal case of highly monochromatic radiation in resonance with an ion transition of very narrow linewidth. In this regime the radiative interaction drives Rabi flopping between the ion internal states (spin states) \cite{Allen-Eberly-two-level-atom-BOOK, Wineland-Meekhof-expt-issues-ion-QC} according to the Hamiltonian\footnote{We use the conventions of \cite{Wineland-Meekhof-expt-issues-ion-QC} in defining $\Omega$ and $\phi_L$.}

\begin{equation}
\op{H}_\Mit{Rabi} = 2 \hbar \Omega \left[ e^{i (- \delta t + \myvec{k} \cdot \myvec{x} + \phi_L)} \sigma_+ + \hc \right]\\ \label{rabi}
\end{equation}

\noindent where $\sigma_+$ is a Pauli operator on the qubit state and the applied radiation has frequency $\omegaud + \delta$, wavevector $\vec{k}$, Rabi frequency $\Omega$, and phase $\phi_L$ at the position of the ion, $\vec{x}$. For an ion initially prepared in $\ket{\da}$, the final state $\Psi(t)$ after an interaction time $t$ with $\delta = 0$ is given by $\Psi(t) = \cos (\Omega t) \ket{\da} - i e^{i \phi_L} \sin (\Omega t) \ket{\ua}$. By varying the duration of the applied radiation, one observes a sinusoidal oscillation of the ion fluorescence with period $\pi/\Omega$. A radiation pulse that induces a total Rabi angle $\Omega t = \pi$ converts $\ket{\da}$ to $\ket{\ua}$ and vice versa. Such a pulse is called a $\pi$ pulse, and one likewise speaks of $\pi/2$ pulses and so forth. Experiments with optical qubits typically require laser intensities on the order of 100 W $\mbox{cm}^{-2}$ to reach Rabi frequencies of hundreds of kHz, so the laser only needs to have a power of a few mW. \\ \\

As will be seen below, it is convenient to use lasers to manipulate hyperfine as well as optical qubits. Usually one drives two-photon stimulated Raman transitions between hyperfine levels, with the Raman laser beams far detuned from any allowed transition so that no spontaneous photon emission can occur. On two-photon resonance, adiabatic elimination of the excited state gives an effective Hamiltonian that has the same form as the Rabi Hamiltonian, Eq.~(\ref{rabi}) \cite{Wineland-Meekhof-expt-issues-ion-QC, Kielpinski-Wineland-thesis}. To simplify the expression for the Raman Rabi frequency $\Omega_R$, we consider the case where only one atomic excited state contributes to the two-photon transition amplitude. The frequency difference between the Raman beams is much less than the detuning $\Delta$ of the Raman laser beams from the excited state. Writing the resonant Rabi frequencies of the two Raman laser beams as $\Omega_1$ and $\Omega_2$, we find $\Omega_R \equiv 2\Omega_1\Omega_2/\Delta$. By conservation of momentum, we find that $\myvec{k}$ in Eq.~(\ref{rabi}) is replaced by the difference $\myvec{k}_1 - \myvec{k}_2$ between the Raman beam wavevectors. Likewise $\phi_L$ is replaced by the difference $\varphi_1 - \varphi_2$ between the phases $\varphi_1$, $\varphi_2$ of the Raman beam electric fields, evaluated at the position of the ion. The ideal Rabi flopping behavior of optical and hyperfine qubits is identical, but in practice laser frequency noise affects the single-photon coupling of optical qubits much more than it affects stimulated Raman coupling of hyperfine qubits, because any common-mode frequency noise of the Raman laser beams cancels out. At a typical excited state detuning of $\approx 100$ GHz, a laser intensity of about 100 W $\mbox{cm}^{-2}$ is again sufficient for Rabi frequencies of hundreds of kHz. \\ \\

\subsection{Coherent control of ion motion}
\label{sec-spin-motion}

So far we have described initialization of our quantum register, operations on single qubits, and detection of the register state. However, one more ingredient is needed to perform universal quantum logic: a gate that entangles two particles \cite{deutuniv,Lloyd-universal-two-qubit-gate,Barenco-Weinfurter-universal-gate-set}. In all deterministic two-ion gates that have been experimentally demonstrated, one indirectly creates a qubit-qubit interaction by transferring quantum information through a motional mode of the ion crystal. After Doppler cooling, the residual thermal excitation of the ions' quantized motion still has a mean occupation number $\overline{n} \approx 5 - 100$ for typical axial trap frequencies of 0.1 - 1 MHz. High-fidelity quantum logic operations require further cooling of the ion motion to nearly $\overline{n} = 0$, which is generally accomplished by resolved-sideband cooling \cite{Wineland-Itano-laser-cooling-theory, Diedrich-Wineland-sideband-cooling, Monroe-Wineland-raman-cooling}, although other methods have been demonstrated \cite{Roos-Blatt-EIT-cooling}. \\ \\

Resolved-sideband cooling operates on the qubit transition of Fig.~\ref{levelschem}, rather than the detection transition. Because the qubit transition is so long-lived, all the motional modes of the ion crystal have frequencies much larger than the transition linewidth. In this case, the ion motion along the laser beam gives rise to sideband transitions offset from the $\ket{\da}$ - $\ket{\ua}$ transition frequency $\omegaud$ by multiples of the trap frequencies, as can be seen from Eq.~(\ref{rabi}). The number of observed sidebands increases with ion temperature, as expected classically. In the following, we need only consider the first-order ``red'' and ``blue'' sidebands of each motional mode, which occur at frequencies $\omegaud - \nu$ and $\omegaud + \nu$ for a motional mode of frequency $\nu$. \\ \\

Red sideband excitation, combined with the initialisation process (see Fig.~\ref{levelschem}), provides a convenient mechanism for cooling to the ground state. When the ion crystal makes a red sideband transition, it loses one phonon in the corresponding motional mode. Initialisation heats the crystal by an amount corresponding to the recoil energy of a single photon, but the recoil energy is much smaller than the energy splitting of motional levels, so the motional mode readily reaches the quantum ground state. Single ions have been cooled to the ground state of motion along the trap axis with residual thermal excitation $\langle n \rangle$ of 0.001 quanta \cite{Roos-Blatt-optical-qubit-control}; in other words, the probability of finding the ion outside the motional ground state was 0.1\%. Three-ion crystals have been cooled to the axial ground state with 99\% efficiency \cite{Barrett-Wineland-teleportation} by successive cooling on the three motional modes. \\ \\

To describe coherent coupling between an ion spin and a motional mode cooled to the ground state, we return to the Rabi Hamiltonian Eq.(~\ref{rabi}), now considering $\myvec{x}$ as the quantum position operator of the ion. If the laser wavevector is parallel to the trap axis, only the sidebands for motion along $\hat{z}$ are observed; many experiments use this geometry and we assume it for simplicity. It is easy to quantize the normal modes of the ion motion, since each mode is just a simple harmonic oscillator. Writing the operator for small displacements of the $i$th ion as $\op{x}_i$ and the conjugate momentum as $\op{p}_i$, we define the annihilation operator $\op{a}_k$ for the $k$th mode of $N$ ions in the usual way:

\begin{equation}
\op{a}_k = \sqrt{\frac{2 \pi N m \nu_k}{2}} \sum_{i=1}^N v_k^{(i)} \left( \op{x}_i + \frac{i}{2 \pi N m \nu_k} \op{p}_i \right) \label{multaop}
\end{equation}

\noindent where $\nu_k$ is the mode frequency, $\myvec{v}_k$ is the normalized eigenvector of ion motional amplitudes for the mode, and $m$ is the ion mass. The dynamics is readily expressed in terms of the Lamb-Dicke parameter $\eta \equiv k_z z_0$, where $z_0 = (2 m_\Mit{ion} \omega_z)^{-1/2}$ is the zero-point wavepacket spread of a single ion along the trap axis and $\myvec{k}$ is the laser wavevector. In the multi-ion case, generalised Lamb-Dicke parameters apply to the various normal modes of the ion crystal; in particular, the center-of-mass mode for $N$ ions always has $\eta_\Mit{COM} = \eta/\sqrt{N}$. For low-error quantum logic, one must arrange that $\eta_k \ll 1$ and $\langle n_k \rangle \eta_k \ll 1$ for all axial motional modes, where the modes are labeled by $k$ and $n_k$ is the number operator of the $i$th mode.\\ \\

The logic lasers in ion-trap QIP experiments are either focused onto one ion at a time, or they illuminate all ions equally. In the former case, the Rabi Hamiltonian Eq.~(\ref{rabi}) near a spin-motion resonance can be approximated as \cite{Kielpinski-Wineland-thesis}

\begin{alignat}{3}
\op{H} & = e^{i \phi_L} \Omega \: \op{\sigma}_{j,+} + \hc & \delta & = \omegaud & \mbox{carrier}\\
& = \eta_k e^{i \phi_L} \Omega \: \op{\sigma}_{j,+} \op{a}_k + \hc & \delta & = \omegaud - \nu_k & \mbox{red sideband} \\
& = \eta_k e^{i \phi_L} \Omega \: \op{\sigma}_{j,+} \op{a}_k^\dagger + \hc & \delta & = \omegaud + \nu_k & \mbox{blue sideband}
\label{motionrabi}
\end{alignat}

\noindent where $\op{a}_k$ is the annihilation operator of the $k$th normal mode and $\myvec{\op{\sigma}}_j$ is the Pauli spin operator of the $j$th ion. For the case of equal illumination, the individual spin operator $\sigma_j$ is replaced by the collective spin operator $\myvec{\op{J}} \equiv \sum_j e^{i \phi_{L,j}} \myvec{\sigma_j}$. The collective spin is just the coherent sum of the individual ion spin operators $\myvec{\sigma}_j$. \\ \\

\begin{figure}
\begin{center}
\includegraphics*[width=8.3cm]{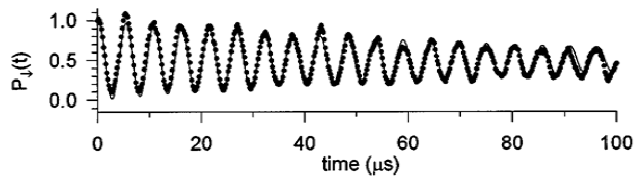}
\caption{Experimentally measured Rabi oscillations on the blue sideband as a function of laser pulse duration, for the initial state $\ket{\da}\ket{n=0}$. After \cite{Meekhof-Wineland-nonclassical-motional-states}.}
\label{rabiosc}
\end{center}
\end{figure}

From Eq.~(\ref{motionrabi}), we see that coherent driving on a motional sideband induces Rabi flopping dynamics between the collective spin state and a particular motional mode. The sideband Hamiltonian is that of the Jaynes-Cummings model, familiar from quantum optics. By driving on the blue sideband of a single ion, one coherently couples the states $\ket{\da}\ket{n}$ and $\ket{\ua}\ket{n+1}$, as demonstrated experimentally in \cite{Meekhof-Wineland-nonclassical-motional-states} (see Fig.~\ref{rabiosc}). A $\pi/2$ pulse on the blue sideband transforms $\ket{\da}\ket{0}$ to the superposition state $\ket{\da}\ket{0} + \ket{\ua}\ket{1}$. (Hereafter we omit normalisation factors for wavefunctions.) This state exhibits entanglement between spin and motion and the $\pi/2$ pulse can be considered as a quantum controlled-NOT gate between the spin qubit and a motional qubit consisting of the states $\{ \ket{n = 0}, \ket{1} \}$ \cite{Monroe-Wineland-spin-motion-XOR}. Many interesting QIP tasks can be performed using only the spin/motion quantum states of a single ion, but these have been thoroughly described elsewhere \cite{Leibfried-Wineland-single-ion-quantum-rev}. In this review, we concentrate on the use of the spin/motion coupling as a means to create spin/spin logic gates. \\ \\

In dealing with hyperfine qubits, one must remember that a Raman transition has an effective wavevector $\deltak = \myvec{k}_1 - \myvec{k}_2$ that is the difference of two optical wavevectors. The effective wavelength for a Raman transition is usually in the optical range, although the transition frequency is in the microwave range. The effective wavevector $\deltak$ only becomes small if the Raman laser beams are nearly copropagating. A common geometry for operations on hyperfine qubits uses Raman beams with an angular separation of $90^\circ$ and with $\deltak$ parallel to the trap axis $\hat{z}$. Then the Raman process only couples to motion along the $\hat{z}$ axis, leading to the simplification of the sideband spectrum noted above. \\ \\

\subsection{Universal quantum logic and QIP protocols}
\label{sec-smallQIP}

A wide variety of two-ion gates can be implemented using the collective motion, from a controlled-NOT \cite{Cirac-Zoller-ion-CNOT, SchmidtKaler-Blatt-CZ-gate} to a controlled phase gate \cite{Milburn-James-phase-gate, Leibfried-Wineland-geometric-gate}, or any of a class of other two-qubit gates \cite{Sorenson-Molmer-bichromatic-gate-PRL, Sackett-Monroe-four-ion-entanglement, Haljan-Monroe-phase-stable-gate}. These gates are deterministic, i.e., they execute a desired unitary operation at a desired moment (apart from technical limitations). Although deterministic gates are not absolutely necessary for large-scale QIP \cite{Knill-Milburn-linear-optics-QC}, access to deterministic gates is a key advantage of ion-trap quantum computing in the near term and drastically reduces the physical resources required to construct a large ion-trap QIP device. \\ \\

Here is a simple example of a quantum gate that uses the shared motion to entangle two ions, along the lines of Cirac and Zoller's original proposal \cite{Cirac-Zoller-ion-CNOT}. This example requires the presence of an additional qubit transition from $\ket{\da}$ to an auxiliary state $\ket{e}$. Suppose we use a $\pi/2$ carrier pulse on both ions to prepare the state $(\ket{\da_1} + \ket{\ua_1}) (\ket{\da_2} + \ket{\ua_2}) \ket{n_k = 0}$, where $n_k$ is the number operator on the $k$th motional mode. The gate then proceeds as follows:

\begin{enumerate}
\item One performs a $\pi$ pulse on the red sideband of ion 1. The red sideband has no effect on $\ket{\da_1} \ket{0}$, so one obtains the state $(\ket{\da_1} \ket{0} + \ket{\da_1} \ket{1}) (\ket{\da_2} + \ket{\ua_2})$.
\item One performs a $2\pi$ pulse on the red sideband of ion 2, but on the auxiliary $\ket{\da}$ - $\ket{e}$ transition. Only the $\ket{\da_2}\ket{1}$ state is affected by this pulse, and it is transformed to $-\ket{\da_2}\ket{1}$ by the action of $H_\Mit{Rabi}$ (Eq.~\ref{rabi}). The resulting state is $\ket{\da_1\da_2}\ket{0} + \ket{\da_1\ua_2}\ket{0} - \ket{\da_1\da_2}\ket{1} + \ket{\da_1\ua_2}\ket{1}$.
\item One performs another $\pi$ pulse on the red sideband of ion 1. Again $\ket{\da_1}\ket{0}$ is unaffected, so the state becomes $\ket{\da_1}(\ket{\da_2} + \ket{\ua_2}) + \ket{\ua_1}(\ket{\da_2} - \ket{\ua_2})$. We can see that this state is entangled by reversing the $\pi/2$ pulse on the carrier, but only on ion 2; then we obtain $\ket{\da_1\ua_2} - \ket{\ua_1\da_2}$, the Bohm-EPR state \cite{Einstein-Rosen-epr,Bohm-quantum-theory-BOOK}.
\end{enumerate}

A more widely used class of two-qubit gates is based on the geometric phase (Berry's phase) associated with transporting the two-ion state around a closed path in the phase space of the motional mode used for logic \cite{Milburn-James-phase-gate, Wang-Molmer-multiqubit-gate}. For instance, simultaneous application of the sideband Hamiltonians of form $\op{J}_x \op{a}_k$, $\op{J}_x \op{a}^{\dagger}_k$ produces a unitary evolution driven by the commutator $\op{J}^2_x$. Such gates have been used in experiments to realize the entangling operators $\sigma_x \sigma_x$ \cite{Sackett-Monroe-four-ion-entanglement} and $\sigma_z \sigma_z$ \cite{Leibfried-Wineland-geometric-gate}. They do not require individual laser addressing of the ions and so they are widely used in QIP experiments. They are also relatively insensitive to the motional state, an important advantage for robust quantum computing experiments because of both imperfect initialisation and ion heating (see Sec.~\ref{sec-decoh}). In some cases one can generalise the two-qubit gate to provide a direct multiqubit interaction \cite{Molmer-Sorensen-multiparticle-entangling-gate, Wang-Molmer-multiqubit-gate}, and up to six ions have been entangled in this way \cite{Leibfried-Wineland-six-atom-entanglement}. In a more algorithmic approach, entangled states of up to eight ions have been created using a sequence of controlled-NOT operations \cite{Haffner-Blatt-scalable-entanglement}. \\ \\

By combining vibrational multiqubit gates with the detection and single-qubit logic techniques outlined above, it is currently possible to construct an ion-trap quantum computer with up to four qubits and capable of implementing tens of logic gates in a computation. Many quantum computing algorithms have been demonstrated using trapped ions, including the Deutsch-Josza algorithm \cite{Gulde-Blatt-deutsch-josza}, the Grover search algorithm \cite{Brickman-Monroe-ion-grover-search}, and the semiclassical Fourier transform \cite{Chiaverini-Wineland-quantum-fourier-xform}, which is the engine of Shor's efficient factoring algorithm \cite{Shor-factoring-algorithm}. Likewise, several quantum communication protocols have been implemented, among them teleportation \cite{Barrett-Wineland-teleportation, Riebe-Blatt-teleportation}, dense coding \cite{Schaetz-Wineland-dense-coding}, and entanglement purification \cite{Reichle-Wineland-entanglement-purification}. \\ \\

These experiments meet the most stringent requirements for QIP with small numbers of qubits. Conceptually, the system is nearly ideal. Every qubit and all relevant motional modes are initialised with low error. Quantum gates with small errors can be performed on demand, and the long qubit coherence time allows the application of a complex sequence of gates. The final state is easy to read out with low error. The quantum states generated in these experiments clearly show entanglement, ruling out classical explanations for the behavior of the quantum register. \\ \\

The QIP protocols that have been demonstrated with trapped ions by no means exhaust the theoretical proposals that can be realised in this system. Many fascinating phenomena of quantum mechanics remain to be elucidated with small ion-trap quantum registers, and alternate approaches to QIP, such as one-way quantum computing \cite{Raussendorf-Briegel-one-way-QC}, can also be explored with trapped ions. Quantum simulations of physical systems, especially in quantum field theory, are an attractive near-term goal for ion-trap QIP. At the same time, it is generally admitted that a new technical approach is needed to scale up ion-trap quantum registers to hundreds of ions and thousands of logic gates. Large ion crystals exhibit a dense sideband spectrum, and unwanted sideband excitations degrade logic gate performance unacceptably for more than $\sim 10$ ions. Most experimental efforts toward large-scale ion QIP now use an array of interconnected ion traps, in a scheme discussed below (Sec.~\ref{sec-qccd}). \\ \\

\section{Errors and error handling in ion-trap QIP}

Large-scale QIP requires long-term storage of quantum information and low-error quantum gates. Errors in classical digital computing are relatively easy to detect and correct by simultaneously processing many copies of the same information. In principle, each logical bit in a classical CPU can be stored in the position state of a single electron, but the CPU actually uses a macroscopic number of electrons, all of which occupy the same state, to store each bit. Implicitly, classical computing relies on the availability of many copies of the same state. However, the destructive nature of quantum measurement makes it impossible to reliably copy a {\it single} instance of a quantum state \cite{Wootters-Zurek-no-cloning}, requiring a radically different approach to error handling. The sensitivity of QIP to physical errors is currently addressed by 1) purely QIP-based error-handling protocols that do not depend on the qubit implementation, 2) careful engineering of qubit encoding and gate operations to take advantage of the qubit physics, and 3) brute-force reduction of sources of technical noise. \\ \\

In order to describe the effect of errors mathematically, the density matrix $\op{\rho}$ supplants the wavefunction as the description of the quantum state of the system, so as to include both quantum coherence and classical (incoherent) randomness on an equal footing \cite{Sakurai-quantum-BOOK}. In a rigorous formalism, the Hamiltonians and unitary operators of fully coherent quantum dynamics are replaced by superoperators \cite{Walls-Milburn-quantum-optics-BOOK}. However, one can often model the effect of a noise source on QIP more simply by assuming that a classical control parameter, like laser power or trap frequency, is varying in a random fashion, and taking an appropriate ensemble average over the realisations of the random cases \cite{Wineland-Meekhof-expt-issues-ion-QC}. \\ \\

\subsection{QIP error handling}
\label{sec-errorhandle}

QIP-based error handling, and especially quantum error correction (QEC) \cite{Shor-QEC, Steane-QEC}, is widely believed to be crucial for the operation of large-scale quantum computers. Without error handling, QIP would consist simply of controlling the quantum evolution of a system so that the initial state encoded a problem, and the final state, a solution. However, complex physical systems usually display chaotic dynamics, i.e., an exponential sensitivity to errors in the initial conditions, so that the maximum length of an accurate computation scales as the log of the error rate. In both classical computing and QIP, frequent and repeated error correction allows accurate computation to proceed indefinitely. This property, called fault tolerance, requires gate and memory errors to remain below certain threshold values; otherwise the process of repeated error correction induces more errors than it repairs. For particular error-handling architectures with a high computational overhead, theoretical estimates of the threshold error can range as high as a few percent \cite{Knill-three-percent-error-correction}, but gate errors below $10^{-3}$ appear computationally desirable, though technically challenging. A basic QEC protocol has been demonstrated with trapped ions, the detection and correction of an ion spin-flip \cite{Chiaverini-Wineland-QEC}, but the gate operations involved in QEC did not reach fault-tolerance, so repeated QEC was not possible. Fault-tolerance necessarily requires an attack on the physical sources of decoherence (Sec.~\ref{sec-physerror}). \\ \\

Decoherence-free subspace (DFS) encoding, a useful complement to QEC, arises in the context of collective decoherence processes, i.e., those processes which have the same effect on each qubit. One of the most prominent decoherence mechanisms in ion-trap QIP is the collective dephasing caused by fluctuating magnetic fields. Since the ion string is very small compared to the spatial wavelength of these magnetic fields, the interaction Hamiltonian is just proportional to $\op{J}_z$. All states of the quantum register with the same number of spins in $\ket{\ua}$ are degenerate with each other. Such a degenerate subspace is called a decoherence-free subspace (DFS) \cite{Zanardi-Rasetti-DFS,Duan-Guo-DFS,Lidar-Whaley-DFS}. Any superposition of states in this DFS is protected from the collective dephasing. The method has been demonstrated to improve memory lifetime in ion-trap QIP under ambient conditions \cite{Kielpinski-Wineland-DFS}. The $\op{J}_x^2$ multiqubit gate, available in ion-trap QIP, enables universal quantum computation on qubits encoded in the dephasing DFS \cite{Kielpinski-Wineland-QCCD}. The DFS concept can be generalised to protection against collective amplitude noise \cite{Zanardi-Rasetti-DFS, Duan-Guo-general-DFS, Lidar-Whaley-DFS}, but this kind of noise does not figure prominently in current ion-trap QIP experiments. \\ \\

\subsection{Diagnostics}
\label{sec-diagnostics}

To apply any error-reduction method effectively, we must clearly understand the methods of quantum state characterisation and their application in ion-trap QIP. A simple diagnostic for single-qubit logic gates is simply to observe the Rabi oscillations of the initial state $\ket{\da}$ under increasing laser pulse duration (see Fig.~\ref{rabiosc}). The slow decay of the oscillation amplitude permits an estimate of the maximum feasible number of single-qubit logic gates, while the form of the decay envelope can help to discriminate between various sources of technical noise \cite{Wineland-Meekhof-expt-issues-ion-QC, Kielpinski-Wineland-thesis}. To sample memory errors independently from gate errors, one implements Ramsey interferometry \cite{Ramsey-molecular-beams-BOOK} by applying a carrier $\pi/2$ pulse, waiting a time $T$, and applying another carrier $\pi/2$ pulse with a variable phase. The probability of finding $\ket{\da}$ oscillates sinusoidally with the phase, and the contrast of the oscillations as a function of $T$ measures the overall error of the Ramsey sequence. Generally $T$ is much longer than the $\pi/2$ pulse duration, so the loss of contrast with increasing $T$ can be attributed purely to memory error. \\ \\

Quantum state tomography is the general QIP method for characterisation of quantum states \cite{Nielsen-Chuang-QIP-book}. In this method, one measures the quantum state in several noncommuting measurement bases in order to reconstruct the density matrix. For a single qubit, one can obtain a complete reconstruction by measuring the spin state along $\sigma_x$, $\sigma_y$, and $\sigma_z$. In ion-trap experiments, one implements the measurement along $\sigma_x$ by performing a $\pi/2$ pulse with $\phi=0$ and subsequently measuring in the usual $\sigma_z$ basis; a similar procedure gives the measurement along $\sigma_y$. Density matrices for experimentally produced two-qubit states can be readily obtained in this manner \cite{Roos-Blatt-bell-state-tomography}. Quantum process tomography extends the tomographic idea to reconstruction of quantum gates. The superoperator corresponding to an experimental application of a gate is fully described by the action of the gate on the qubit basis states. For a two-ion gate, one performs quantum state tomography of the final state produced from $\ket{\da\da}$, then for that produced from $\ket{\da\ua}$, and so on. A second tomographic reconstruction from the final-state density matrices gives the gate superoperator. The diagnostic use of process tomography has yielded experimental improvements for two-ion entangling gate implementations and also enables comparisons between various gate schemes \cite{Riebe-Blatt-process-tomography}. \\ \\

\begin{figure}
\begin{center}
\includegraphics*[width=8.3cm]{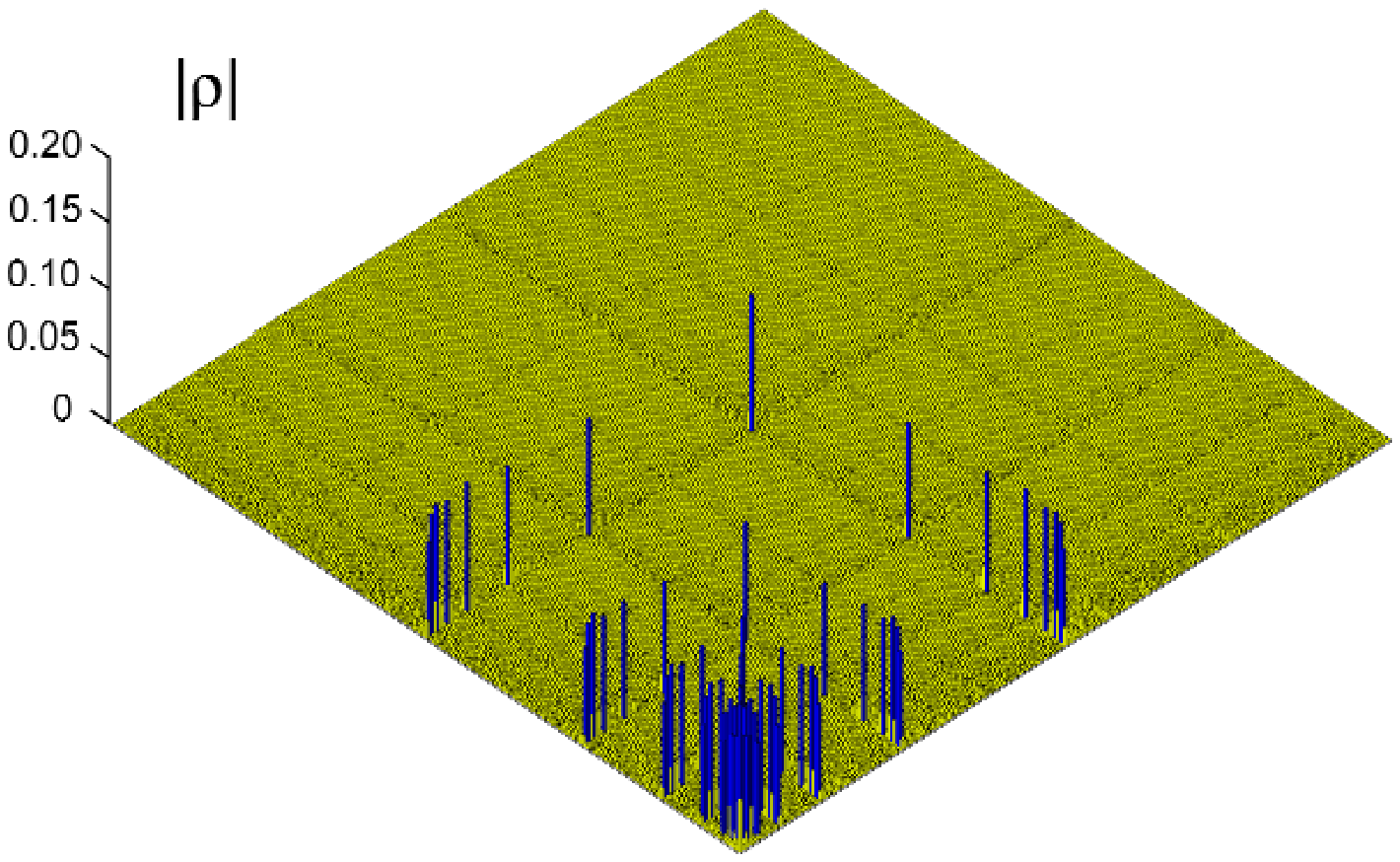} 
\caption{Partial data from a tomographic analysis of eight ion qubits prepared in the W-state $\ket{\ua\da\ldots\da} + \ket{\da\ua\ldots\da} + \ldots + \ket{\da\da\ldots\ua}$ \cite{Haffner-Blatt-scalable-entanglement}. Only the absolute value of the density matrix elements is shown. Corresponding data on the phase of the density matrix elements (not shown) completes the tomography dataset. Figure courtesy of R. Blatt and H. H{\"a}ffner, U. Innsbruck.}
\label{tomography}
\end{center}
\end{figure}

Unfortunately, quantum state tomography is resource-intensive. Measuring the state populations along $\sigma_z$ to 1\% accuracy requires on the order of $10^4$ experimental repetitions, even for quantum-limited detection. Tomography of a general $N$-particle state requires measurement along ${\mathcal O}(n^2)$ noncommuting measurement bases. For eight ions, this poses a significant challenge \cite{Haffner-Blatt-scalable-entanglement}, as illustrated in Fig.~\ref{tomography} . Quantum process tomography requires even more resources, scaling as ${\mathcal O}(n^4)$ for an $n$-qubit gate. Although theoretical efforts are underway to simplify tomography of large quantum systems, many other diagnostic methods currently supplement tomographic analysis in ion-trap QIP. \\ \\

Multi-qubit states often lend themselves to a partial, but highly informative, characterisation by specially designed measurements.
A classic benchmark for verifying ``quantumness'' of a two-qubit state is the violation of a Bell inequality \cite{Bell-inequality}. Bell inequalities give upper bounds on classical two-particle correlation functions, but quantum mechanics predicts that higher correlation values should be possible, and indeed they are observed in experiments on two ions \cite{Rowe-Wineland-Bell-inequality}. Because the assumptions required to derive a Bell inequality are very general, experimental tests of Bell inequalities show that the counterintuitive features of quantum mechanics are essential for a correct physical theory. \\ \\

In some cases, a special measurement is scalable in that it remains useful for entangled states involving more and more qubits. For instance, the state $\ket{\da\da \ldots \da} + \ket{\ua\ua \ldots \ua}$ has been recently produced for up to six ions \cite{Leibfried-Wineland-six-atom-entanglement}. Ideally, the density matrix has only four nonzero elements: $\rho_{\dan,\dan}$, $\rho_{\uan,\uan}$, $\rho_{\dan,\uan}$, and $\rho_{\uan,\dan} = \rho^*_{\dan,\uan}$. The diagonal population elements $\rho_{\dan,\dan}$, $\rho_{\uan,\uan}$ are readily measured by the usual detection methods. To measure the experimental value of the coherence $\rho_{\dan,\uan}$, one applies a carrier $\pi/2$ pulse simultaneously to all ions with a variable phase $\phi$. One then detects the number $N_\da$ of ions in $\ket{\da}$ and computes the parity $\Pi(\phi) \equiv (-1)^{N_\da}$. Assuming an ideal carrier pulse, one finds

\begin{equation}
\Pi(\phi) = 2 | \rho_{\dan,\uan} | \cos N \phi \label{parosc}
\end{equation}

\noindent so one extracts the desired coherence element of the density matrix by measuring the parity as a function of phase. The deviation of $|\rho_{\dan,\uan}|$ from its ideal value of 0.5 provides useful bounds on a wide variety of error processes. One can also place limits on other sources of gate error by Fourier analysis of $\Pi(\phi)$. \\ \\

\subsection{Physical errors}
\label{sec-physerror}

The physical error sources during quantum logic operations are quite different from the sources of ``memory'' errors incurred during the times when an ion is sitting idle. In general, logic errors dominate, and their sources are quite different for optical than for hyperfine qubits, while the sources of memory errors are common to both kinds of qubit. The minimum gate error reported in the refereed literature is 3\%, obtained with a geometric phase gate operating on a Raman transition between hyperfine states \cite{Leibfried-Wineland-geometric-gate}. For a hyperfine qubit, spontaneous emission from the excited state in the Raman process dominates the gate error \cite{Leibfried-Wineland-geometric-gate}. Spontaneous emission error can be reduced to an arbitrarily low value by increasing the excited-state detuning, at the cost of increased Raman laser power \cite{Ozeri-Wineland-Rayleigh-scattering-decoherence}. Theoretical calculations for a number of ion species show that laser intensities of perhaps $10^5\: \mbox{W cm}^{-2}$ and detunings of 10 THz are required for error rates on the order of $10^{-3}$ \cite{Ozeri-Wineland-spontaneous-emission-gate-error}. For an optical qubit, the dominant source of gate error is usually laser frequency noise \cite{SchmidtKaler-Blatt-CZ-gate}, which is purely technical in origin. Even so, an ultrastable laser linewidth on the order of 100 Hz can contribute as much as 10\% error \cite{Roos-Blatt-bell-state-tomography}. A recent preprint from the Innsbruck group gives an error of $7 \times 10^{-3}$, obtained by a geometric phase gate operating on an optical qubit \cite{Benhelm-Blatt-low-gate-error}, not impossibly far from the target error rate of $10^{-4}$ commonly assumed for large-scale ion-trap QIP architectures \cite{Kim-Slusher-ion-trap-system-design, Steane-large-scale-ion-QC}. \\ \\

\begin{figure}
\includegraphics*[width=8.3cm]{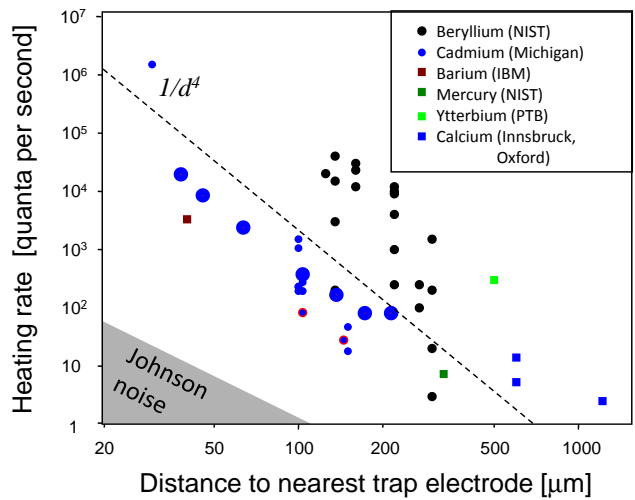}
\caption{Heating rate measurements for ion traps of different sizes $d$, expressed in units of motional quanta per second. The measurements were taken over the last ten years by researchers around the world, using a variety of macroscopic and microfabricated traps, and follow a general $d^{-4}$ power law. The large blue dots represent data taken with a single trap with movable electrodes, and clearly rule out the $d^{-2}$ scaling expected for Johnson (thermal) noise \cite{Deslauriers-Monroe-ion-heating-scaling}. The large scatter of the data is only partly understood. Figure courtesy of C. Monroe, U. of Maryland.}
\label{heating}
\end{figure}

Although heating of the ion motion is not a major error source during a single gate operation, recooling the ions would require spontaneous scattering of laser light and might easily destroy the quantum coherence. In current QIP experiments, the thermal excitation continues to accumulate and can degrade the two-qubit logic gates toward the end of a long gate sequence, so significant efforts have been expended to understand and eliminate ion heating. Unfortunately, the exact mechanism of ion heating remains elusive, although some features of the experimental data can be understood phenomenologically. The heating almost certainly arises from fluctuating electric fields that are nearly spatially uniform over the extent of a few-ion crystal \cite{James-ion-heating-theory, King-Wineland-two-ion-raman-cooling}. However, the observed heating rates indicate fluctuating fields that are much stronger than those expected from thermal noise \cite{Turchette-Wineland-ion-heating}, and the inferred spectral density of electric field noise scales approximately as $1/f$, rather than the constant spectral density of thermal noise. As shown in Fig.~\ref{heating}, the heating rate scales with trap size $d$ as approximately $d^{-4}$; this result has been rigorously confirmed using a trap with movable electrodes \cite{Deslauriers-Monroe-ion-heating-scaling}. These observations are consistent with the idea that the heating is driven by the movement of small patches of charge on the trap electrodes, perhaps caused by adsorbed gas \cite{Wineland-Meekhof-expt-issues-ion-QC, Turchette-Wineland-ion-heating}. It also appears that contamination of the electrodes by the atomic beam used for loading can substantially increase the heating rate \cite{Turchette-Wineland-ion-heating, DeVoe-Kurtsiefer-ion-heating-trap-instability, Rowe-Wineland-ion-transport}, perhaps accounting for some of the scatter of the data in Fig.~\ref{heating}. Recent results show that the ion heating is suppressed by several orders of magnitude at cryogenic temperatures \cite{Deslauriers-Monroe-cryogenic-ion-heating, Labaziewicz-Chuang-cryogenically-suppressed-heating}, though the thermal noise limit has not yet been observed. \\ \\

Memory errors are well understood in trapped-ion QIP. The dephasing of a two-level atomic system is the fundamental limitation to the precision of atomic clocks. The efforts of the metrology community over the last few decades have already uncovered, measured, and found ways to minimise the sources of decoherence that apply to an ion-trap quantum memory, and these methods are rapidly being adapted for trapped-ion QIP. The error rates for memory implementations depend on the precise choice of ion energy levels for representing the qubit states. A fundamental limit to the memory lifetime of an optical qubit is given by the radiative decay rate, of order 1 second in current experiments, while hyperfine qubits have radiative lifetimes comparable to the age of the universe. \\ \\

The dominant source of memory error in current experiments is generally dephasing by ambient magnetic field noise, which shifts the qubit energy levels through the Zeeman effect. Even when the quantum information is encoded in a DFS, the residual error caused by fluctuating magnetic field {\it gradients} limits memory time \cite{Haffner-Blatt-robust-entanglement}. Magnetic field dephasing can be suppressed by several orders of magnitude by choosing a qubit transition with vanishing first-order Zeeman shift, a fact well known by the metrology community and explicitly demonstrated in the QIP context \cite{Langer-Wineland-long-lived-memory}. A two-ion gate compatible with such a ``clock'' transition has been demonstrated \cite{Haljan-Monroe-entangled-clock-state} and has been found to be resistant to laser phase fluctuations \cite{Haljan-Monroe-phase-stable-gate}. A DFS made with clock-state qubits might offer entangled states with lifetimes far in excess of the current record of 20 seconds \cite{Haffner-Blatt-robust-entanglement} and on to the 10 minute coherence times demonstrated for ion-trap atomic clocks \cite{Bollinger-Wineland-10-minute-coherence}. \\ \\

\section{Toward large-scale ion trap quantum computing}
\label{sec-largescale}

\subsection{The QCCD architecture}
\label{sec-qccd}

As we have seen, one can use a small number of trapped ions to construct a quantum register. However, manipulating a large number of ions in a single trap presents immense technical difficulties, and scaling arguments suggest that a single trap can only support computations on tens of ions \cite{Hughes-Petschek-ion-QIP-decoherence-bounds,Wineland-Meekhof-expt-issues-ion-QC,daphnascale}. To build up a large-scale quantum computer, Wineland and co-workers proposed a ``quantum charge-coupled device" (QCCD) architecture consisting of a large number of interconnected ion traps \cite{Wineland-Meekhof-expt-issues-ion-QC, Kielpinski-Wineland-QCCD}. By adjusting the operating voltages of these traps, one can confine a few ions in each trap or shuttle ions from trap to trap.\footnote{The QCCD architecture is named in analogy to the ubiquitous charge-coupled device (CCD) camera, which uses shuttling of electrons for image readout.} In any particular trap, one can manipulate a few ions using the methods already demonstrated, while the connections between traps allow communication between sets of ions. Since both the speed of quantum logic gates \cite{Steane-Blatt-ion-QC-speed} and the shuttling speed are limited by the trap strength, shuttling ions between memory and interaction regions can consume an acceptably small fraction of a clock cycle. The QCCD architecture is not only a dream: the NIST group used small QCCD devices to demonstrate many of the protocols discussed in Sec.~\ref{sec-smallQIP} \cite{Barrett-Wineland-teleportation, Chiaverini-Wineland-quantum-fourier-xform, Schaetz-Wineland-dense-coding, Reichle-Wineland-entanglement-purification}. \\ \\

\begin{figure}
\begin{center}
\includegraphics*[width=8.3cm]{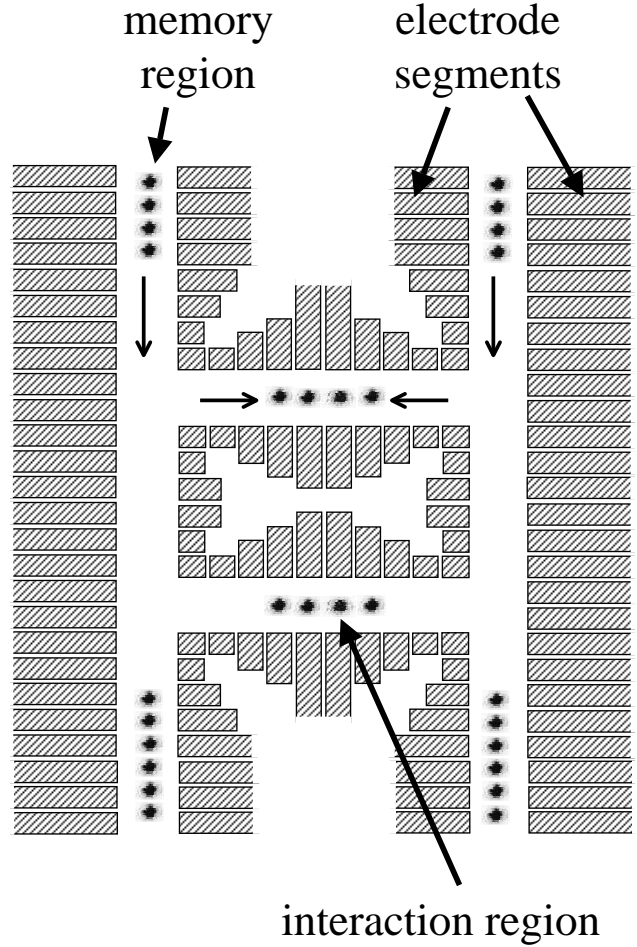}
\caption{Schematic of the QCCD architecture for large-scale ion-trap QIP, after \cite{Kielpinski-Wineland-QCCD}. Ions (dots) are confined to the local trap axis by RF potentials and are shuttled between memory, interaction, and detection regions by quasistatic potentials applied to the DC electrodes (shaded boxes). For simplicity, only the DC electrodes are shown; the RF electrodes can be implemented in a variety of geometries (see text).}
\label{qccd}
\end{center}
\end{figure}

Figure \ref{qccd} shows a schematic of an ion-trap array used in the QCCD architecture. Trapped ions storing quantum information are held in the memory regions. To perform a logic gate, one moves the relevant ions into an interaction region by applying appropriate voltages to the electrode segments. In the interaction region, the ions are held close together, enabling the Coulomb coupling necessary for entangling gates. Lasers are focused through the interaction region to drive the gates. Detection can occur in the interaction regions or in separate parts of the trap array. The QCCD array can be considered as a network of single-trap QIP devices, so that the architecture is essentially modular. The techniques used for quantum logic are those already demonstrated for single-trap quantum registers, so many of the problems of a large-scale QCCD device, for instance the problems of fabricating large trap arrays and of addressing the many interaction regions with laser beams, are easily understood in terms of classical physics and engineering \cite{Kim-Slusher-ion-trap-system-design}. \\ \\

One can create the trapping and transport potentials needed for the QCCD using a combination of radiofrequency (RF) and quasistatic electric fields. Figure \ref{qccd} shows a conceptual picture, including only the electrodes that support the quasistatic fields. By varying the voltages on these electrodes, we can perform ion transport and confinement along the local trap axis, which lies along the arrows in Figure \ref{qccd}. Two more layers of electrodes lie above and below the static electrodes. Applying RF voltage to the outer layers confines the ions transverse to the local trap axis, just as for the standard linear trap. This geometry allows stable transport of the ions around $\mathsf{T}$- and $\mathsf{X}$-junctions, so one can build complex, multiply connected trap structures. \\ \\

The first experiments toward the QCCD scheme used a trap with six axial electrode segments, giving two trapping regions separated by 1.2 mm. Single ions were reliably transported between the trapping regions in $\sim 20 \: \mu$s, with little loss of qubit coherence and motional heating much less than 0.1 quantum per round trip \cite{Rowe-Wineland-ion-transport}. Mastering the more complex task of splitting an ion crystal with low heating \cite{Barrett-Wineland-teleportation} enabled the NIST group to demonstrate some of the three- and four-qubit QIP protocols discussed in Sec.~\ref{sec-qccd}. The extension of these linear trap arrays to $\mathsf{T}$- and $\mathsf{X}$-junctions is challenging because residual RF fields cause unwanted trapping potentials along the ion transport direction. Transport around a $\mathsf{T}$-junction has been achieved with less than 2\% error, at the cost of a massive heating of the ion motion to a few thousand kelvin \cite{Hensinger-Monroe-T-junction-trap}. In principle, these residual RF fields can be eliminated by careful trap design. Recent work indicates that transport speed can be increased by an order of magnitude using careful shaping of the voltage waveforms that drive the trap electrodes \cite{Huber-SchmidtKaler-fast-ion-transport}, but it remains to be seen whether this technique leads to substantial heating from the quantum ground state. \\ \\

Even with the best trap design, it is hard to imagine that the ions will remain in the motional ground state indefinitely, especially under transport. Although laser cooling of the qubits cannot occur during QIP, {\it sympathetic cooling} of the qubit ions by another ion species has been experimentally demonstrated \cite{Blinov-Monroe-sympathetic-cooling, Barrett-Wineland-sympathetic-cooling}. Confining both species in the interaction region allows use of the cooling species as a heat sink, with the Coulomb interaction providing energy transfer from the qubit ions. No decoherence need occur if the laser wavelengths relevant to the cooling species are sufficiently far detuned from the transitions of the qubit species. \\ \\

Additional decoherence mechanisms for the qubit states also arise from ion transport. For instance, the spatial variations of the magnetic field strength along the transport path shift the qubit energy levels through the Zeeman effect, so that, e.g., $\ket{\da} + \ket{\ua} \rightarrow \ket{\da} + e^{i\alpha} \ket{\ua}$ with a phase $\alpha$ depending on the transport path. Current experiments compensate for these phases using spin-echo refocusing, i.e., by swapping the states $\ket{\da}$ and $\ket{\ua}$ in an appropriate way during ion transport.  \\ \\

\subsection{Implementing the QCCD}

The QCCD architecture requires large-scale system integration of widely disparate electronic and optical techniques in an ultra-high vacuum environment. Microfabrication of trap arrays has been discussed above in Sec.~\ref{sec-trap}. Low-noise, rapid ion transport demands parallel delivery of precisely shaped arbitrary voltage waveforms to all trap electrodes, a need currently met by computer-controlled digital-to-analog converters \cite{Hensinger-Monroe-T-junction-trap, Huber-SchmidtKaler-fast-ion-transport}. However, this technology is difficult to scale past a few tens of electrodes and integration of classical control circuitry with the trap array appears essential in the long run \cite{Kim-Slusher-ion-trap-system-design}. Such integration would ideally enable programming of ion transport sequences by simple digital commands from an external computer. A recently demonstrated silicon-based surface trap \cite{Britton-Wineland-silicon-surface-trap}, fabricated using a CMOS-compatible process, represents a first step toward integration of classical circuitry with the QCCD. \\ \\

Highly parallel quantum logic and initialisation operations require that an array of laser beams be delivered to the trap sites through switching or repositioning. Currently, laser addressing is achieved by changing the deflection angle of an acousto-optic modulator \cite{Nagerl-Blatt-individual-addressing}. In recent work, a micromirror array fabricated by a microelectromechanical systems (MEMS) process was used to steer a near-infrared laser beam, allowing laser addressing on a $5 \times 5$ grid of spacing $8 \mu$m with a switching time of $\sim 10 \:\mu$s \cite{Knoernschild-Kim-MEMS-for-QIP}. Similar MEMS micromirrors have been fabricated with high reflectivity at the UV wavelengths relevant to ion QIP \cite{Kim-Kim-MEMS-micromirrors-ion-QC}. These devices operate outside the UHV chamber, allowing wide latitude in the choice of fabrication technology. \\ \\

\begin{figure}
\begin{center}
\includegraphics*[width=8.3cm]{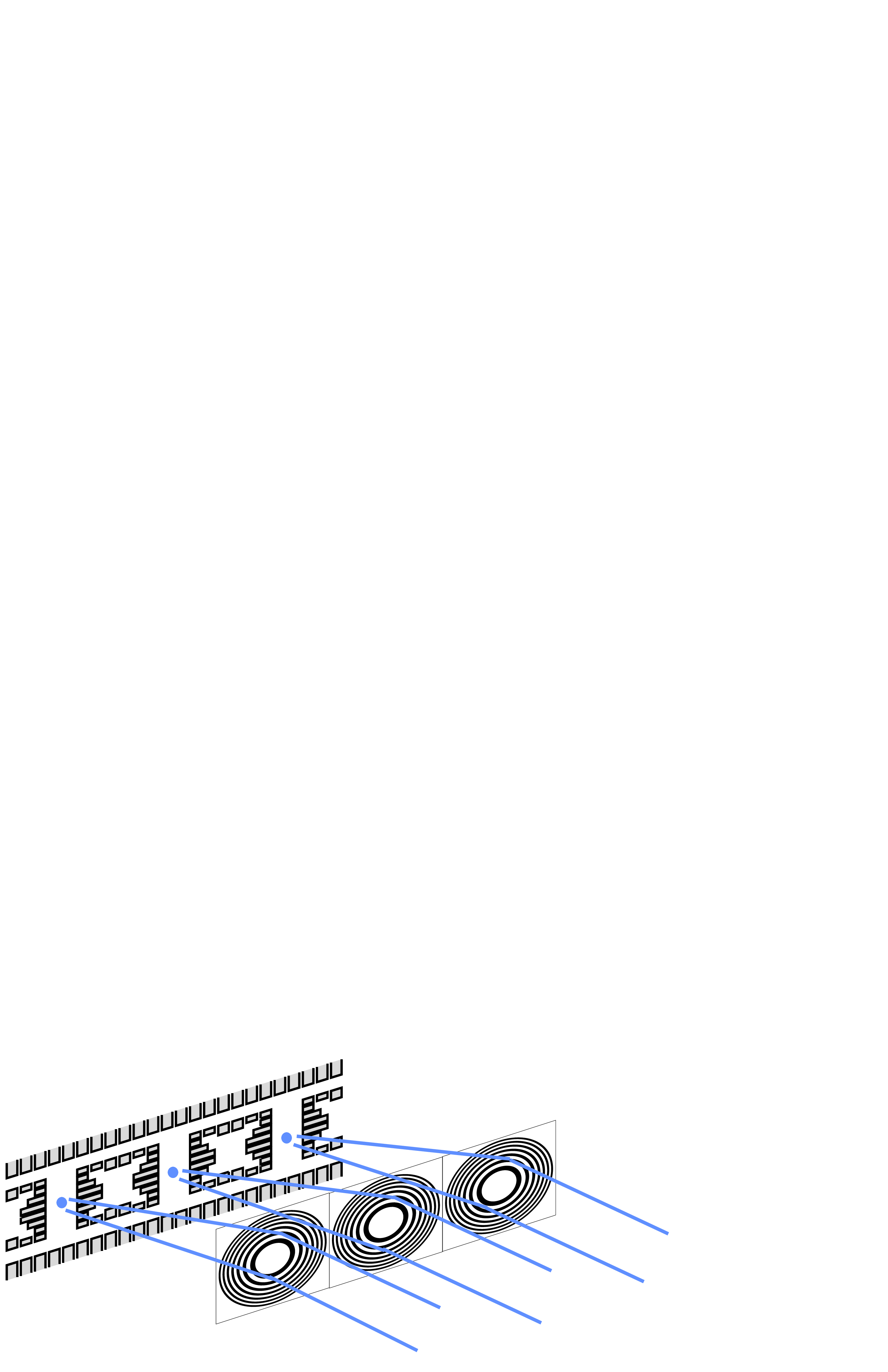}
\caption{Integration of a trap array with a microfabricated array of phase Fresnel lenses. The lenses collimate ion fluorescence from the individual detection zones for reimaging onto detectors or optical fibers.}
\label{fig-pflarray}
\end{center}
\end{figure}

Methods for highly parallel detection of ion fluorescence have not been explored much. The objective lenses used in current experiments have apertures $\sim$ 50 mm and a useful field of view of $< 1$ mm, so only one or two detection regions can be currently realised. The semiconductor industry uses more complex multi-element lenses with field of view $> 10$ mm for lithographic exposures, but these lenses are the result of a massive engineering effort, concentrated at a few specific wavelengths, that would be difficult to replicate in the QIP community. Alternatively, one can realise parallel detection using an array of light collection regions, one for each detection site. The author and co-workers have proposed using a microfabricated array of Fresnel lenses for this purpose \cite{Streed-Kielpinski-fresnel-lens}, as shown in Fig.~\ref{fig-pflarray}. In \cite{Streed-Kielpinski-fresnel-lens} a single phase Fresnel lens with $f = 3$ mm was designed for the 369.5 nm \yb transition and was fabricated by lithography and etching of fused silica, in a scalable and UHV-compatible process. From optical tests, we predict a fluorescence collection efficiency of 4\% for this lens, several times higher than current multi-element lens systems \cite{Moehring-Monroe-separated-ion-entanglement} and nearly sufficient for fault-tolerant QIP \cite{Steane-large-scale-ion-QC}. \\ \\

\subsection{Quantum interfacing of ions and photons}
\label{sec-interface}

Since trapped ions are essentially stationary, quantum communication between ion-trap QIP devices will require the coherent transfer of quantum information between ion qubits and an altogether different physical qubit implementation that is better suited to long-distance travel. Single photons are a natural candidate for this quantum interface, in view of the well-developed QIP techniques in linear optics, and since ion QIP is already heavily reliant on optical technology. Optical quantum communication among ion-trap QIP devices may prove a useful route to large-scale quantum computing \cite{Duan-Monroe-probabilistic-ion-photon-QC}. A near-term application for ion-trap QIP is the construction of a quantum repeater for optical quantum cryptography over long distances \cite{Moehring-Monroe-ion-photon-network-rev}. \\ \\

So far, the most complex ion-photon networking tasks have been performed by probabilistic detection of ion fluorescence. This method relies on the entanglement between the final state of a spontaneously emitting ion and the polarisation of the emitted photon, first demonstrated in \cite{Blinov-Monroe-ion-photon-entanglement}. If the single photons emitted in fluorescence from two widely separated ions are made to interfere at a beamsplitter, a subsequent coincident detection of the photons projects the ions into an entangled state. Remote entanglement of two ions in independent traps separated by a meter has been achieved \cite{Moehring-Monroe-separated-ion-entanglement} and shown to violate a Bell inequality \cite{Matsukevich-Monroe-separated-ion-Bell-test}. \\ \\

\begin{figure}
\begin{center}
\includegraphics*[width=8.3cm]{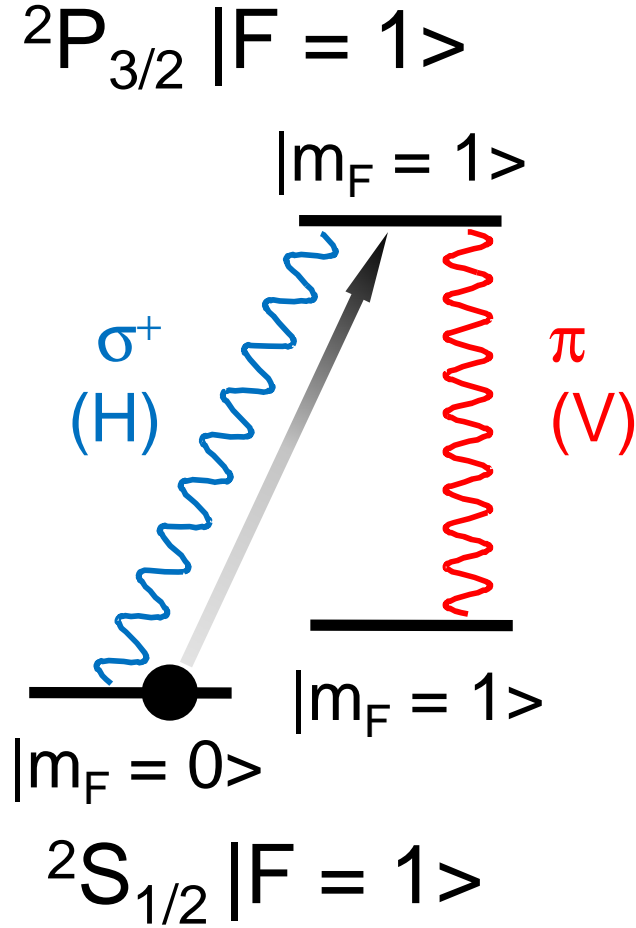}
\caption{Protocol for generating probabilistic ion-photon entanglement used in \cite{Blinov-Monroe-ion-photon-entanglement}. Laser excitation is followed by spontaneous emission of a single photon whose polarisation is correlated with the final atomic state.}
\label{fig-emissionpaths}
\end{center}
\end{figure}

Figure~\ref{fig-emissionpaths} illustrates the protocol for ion-photon entanglement used in \cite{Blinov-Monroe-ion-photon-entanglement}. A single ion with $^{2}{\mbox S}_{1/2}$ ground state and nuclear spin 1/2 is initially prepared in the $\ket{F,m_f}=\ket{1,0}$ hyperfine level. Laser excitation to the $^{2}{\mbox P}_{3/2} \ket{F'=2,m_{F'}=1}$ state is followed by spontaneous emission of a single photon. If the photon has $\pi$ polarisation, the final atomic state must be $\ket{F,m_F}=\ket{1,1}$, while for $\sigma^+$ polarisation the final state must be $\ket{1,0}$. In the plane perpendicular to the quantisation axis, an emitted $\pi$ ($\sigma^+$ photon is vertically (horizontally) polarised. The statistics of ion and photon measurement results under single-qubit rotations reveal the nonclassical correlations. This protocol is readily extended to remote ion-ion entanglement through the nonclassical interference of emitted photons at a beamsplitter \cite{Moehring-Monroe-separated-ion-entanglement}. \\ \\

Several experiments have taken steps toward deterministic coupling between ion and photon quantum states by using high-finesse resonators to collect a large fraction of the photons spontaneously emitted by a single ion. The subwavelength confinement of the ion allows fine control over the ion-photon coupling \cite{Guthohrlein-Walther-ion-cavity-coupling} and selective excitation of the vibrational sidebands that the ion imposes on the cavity mode \cite{Mundt-Blatt-ion-cavity-coupling}. Ion-cavity systems have been used to demonstrate stable probabilistic single-photon generation over tens of minutes \cite{Keller-Walther-ion-cavity-single-photon}, illustrating their suitability as robust building blocks for QIP. However, in current experiments the single-ion cooperativity parameter $C = g^2/(2 \kappa \gamma)$ is on the order of one, so the coherence of the ion-photon coupling is low. Here $g$ is the Rabi frequency of the ion when a single photon is present in the cavity, $\kappa$ is the cavity decay rate, and $\gamma$ is the decay rate of the ion transition. Proximity effects pose a major obstacle to increasing $C$. The short ($\lesssim 100 \mu$m) resonators used for neutral-atom cavity experiments achieve high $C$ largely because of their small mode volume. Since the single-photon energy remains constant, the electric field of a single photon grows, and so does $g$. This path is not available in ion-trap experiments, making high $C$ solely reliant on high mirror reflectivity. \\ \\

\section{Conclusion}

All physical implementations of QIP are currently in their infancy. None has yet clearly demonstrated the capacity for fault-tolerant quantum computing or the possibility of outcompeting classical computers in any way. Nevertheless, ion-trap experiments fulfill the requirements for effective small-scale QIP. Ion qubits can be initialised and read out on demand, and the deterministic quantum logic gates afforded by the motional modes enable universal quantum logic. The low error in these basic QIP operations has allowed successful tests of many simple QIP protocols.\\ \\

Trapped ions appear to be an attractive system for large-scale quantum computing under the QCCD architecture. Experimental progress toward the QCCD has already advanced the state of the art in small-scale QIP. The ion-trap community is successfully addressing the technical problems of scaling, namely trap array fabrication, fast laser beam steering, and parallel detection. Demonstrations of quantum interfacing of ions with photons promise optical quantum communication between QCCD devices. With these advantages, trapped ions will likely occupy a prominent role in QIP for many years to come. \\ \\

\acknowledgments{I thank Rainer Blatt, Wolfgang Lange, Chris Monroe, and David Wineland for helpful discussions. This work was
supported under Australian Research Council grants DP0773354 (Kielpinski) and FF0458313 (Wiseman) and by the
US Air Force under grant FA4869-08-1-4005.}

\bibliography{bib0308}

\end{document}